# Phase Improvement Algorithm for NLFM Waveform Design to Reduction of Sidelobe Level in Autocorrelation Function

R. Ghavamirad, M. A. Sebt and H. Babashah

In this paper, a phase improvement algorithm has been developed to design the nonlinear frequency modulated (NLFM) signal for the four windows of Raised-Cosine, Taylor, Chebyshev, and Kaiser. We have already designed NLFM signal by stationary phase method. The simulation results for the peak sidelobe level of the autocorrelation function in the phase improvement algorithm reveal a significant average decrement of about 5 dB with respect to stationary phase method. Moreover, to evaluate the efficiency of the phase improvement algorithm, minimum error value for each iteration is calculated.

*Introduction:* Goal of pulse compression is to increase bandwidth and improve range resolution [1]. There are several methods for pulse compression. For example, coding methods such as Barker, Huffman, Zadoff-Chu, etc. are utilized in pulse compression [2], but due to the phase discontinuity and the signal amplitude variability (such as the Huffman codes), they result in loss increment in the receiver (due to mismatching) [3]. The linear frequency modulation (LFM) method has received much attention since its phase continuity and the constant amplitude of the signal, but it suffers from relatively high sidelobes in autocorrelation function (ACF) [3].

The NLFM method has been proposed to reduce the sidelobes level in ACF. In NLFM method, the signal amplitude is constant and the frequency variations with respect to time is nonlinear. Stationary phase concept (SPC) is commonly used in NLFM method. SPC explains that power spectral density (PSD) in a frequency is relatively high if the related frequency variation is low with regard to time [3]. Using this method leads to noticeable sidelobes level decrement in ACF. Additionally, it causes the main lobe width to increases slightly but negligible.

The phase improvement algorithm (PIA) is proposed here to be used after the stationary phase method. This method is designed based on the phase matching techniques. To start the algorithm, an appropriate value for the phase is used which comes from stationary phase method. The algorithm is repeated several times in order to get closer to the optimal phase value where sidelobes level are significantly reduced compared to the stationary phase method.

The remainder of the letter is organized as follows: Second section outlines the proposed phase improvement algorithm. In the third section, the simulation results of the proposed algorithm are discussed and a comparison between SPC and the proposed method is made. Finally, the fourth section concludes the paper.

*NLFM Signal Design with Phase Improvement Algorithm:* In the phase improvement algorithm, the goal is to find the desired signal phase in terms of the minimum error. To achieve this goal, first, a window is selected as the initial PSD. Then, the signal phase is obtained at each iteration by reducing the error of the algorithm. Due to the constant amplitude of the NLFM signal, the desired signal can be shown as follow

$$x(t) = A \exp(j\varphi(t)), \qquad |t| \leq \frac{T}{2} \quad (1)$$

where $\varphi(t)$ is the signal phase and $A$ is signal amplitude, which is constant and $T$ is pulse width. If $X(f)$ is Fourier transform of $x(t)$, so

$$X(f) = \int_{-\frac{T}{2}}^{\frac{T}{2}} x(t) \exp(-j2\pi ft) \, dt \quad (2)$$

Suppose $|Y(f)|$ is the root of the initial PSD; therefore, the following equation is used to calculate the difference between the amplitude of $|X(f)|$ and $|Y(f)|$.

$$Error = \int_{-\frac{B}{2}}^{\frac{B}{2}} (|Y(f)| - |X(f)|)^2 \, df \quad (3)$$

where $B$ is the bandwidth of $x(t)$. If the difference between the two complex numbers decreases, it can be concluded that their amplitude difference is also reduced; therefore, if the phase of $X(f)$ is $\theta(f)$, (3) is rewritten as shown in below to be applicable in the phase matching [4,5].

$$Error = \int_{-\frac{B}{2}}^{\frac{B}{2}} ||Y(f)| \exp(j\theta(f)) - X(f)|^2 \, df \quad (4)$$

The primary aim is to obtain $X(f)$ in order to achieve $x(t)$ by taking an inverse Fourier transform [6]. Consider $y_m(f) = |Y(f)| \exp(j\theta(f))$ and $k = (K-1)f/B$; Therefore, we can solve the integral in the discrete form.

$$Error = \sum_{k=0}^{K-1} |y_m(k) - X(k)|^2$$
$$= \sum_{k=0}^{K-1} (y_m(k) - X(k))^* (y_m(k) - X(k)) \quad (5)$$

$$X(k) = \sum_{n=0}^{N-1} x(n) \exp\left(-j\frac{2\pi kn}{K}\right) \quad (6)$$

The symbol $*$ in (5) indicates complex conjunction and $N$ in (6) is the number of samples in signal $x(n)$. Now, consider $\mathbf{y_m} = [y_m(0) \; y_m(1) \; y_m(2) \; ... \; y_m(K-1)]^T$ and $\mathbf{x} = [x(0) \; x(1) \; x(2) \; ... \; x(N-1)]^T$. Thus, we have

$$X(k) = \begin{bmatrix} 1 & \exp\left(-j\frac{2\pi k}{K}\right) & ... & \exp\left(-j\frac{2\pi (N-1)k}{K}\right) \end{bmatrix} . \mathbf{x} \quad (7)$$

Making use of $w^{k,n} = \exp\left(-j\frac{2\pi kn}{K}\right)$, the following equation for $X(k)$ can be expressed in vector space.

$$\begin{bmatrix} X(0) \\ \vdots \\ X(K-1) \end{bmatrix} = \begin{bmatrix} w^{0,0} & \cdots & w^{0,(N-1)} \\ \vdots & \ddots & \vdots \\ w^{(K-1),0} & \cdots & w^{(K-1),(N-1)} \end{bmatrix}_{K \times N} \begin{bmatrix} x(0) \\ \vdots \\ x(N-1) \end{bmatrix} \quad (8)$$

In (8), $\mathbf{W}_{K \times N}$ is the discrete Fourier transform (DFT) matrix. We now take a partial derivative of (5) with respect to vector $\mathbf{x}$.

$$Error = (\mathbf{y_m} - \mathbf{Wx})^H (\mathbf{y_m} - \mathbf{Wx})$$
$$= \mathbf{y_m}^H \mathbf{y_m} - \mathbf{y_m}^H \mathbf{Wx} - \mathbf{x}^H \mathbf{W}^H \mathbf{y_m} + \mathbf{x}^H \mathbf{W}^H \mathbf{Wx}$$
$$\rightarrow \frac{\partial(Error)}{\partial \mathbf{x}} = -(\mathbf{W}^H \mathbf{y_m} - \mathbf{W}^H \mathbf{Wx})^* = 0$$
$$\rightarrow \mathbf{x} = (\mathbf{W}^H \mathbf{W})^{-1} \mathbf{W}^H \mathbf{y_m} \quad (9)$$

where $\mathbf{W}^H$ is Hermitian (complex conjugate transpose) of $\mathbf{W}$; $\mathbf{y_m}$ is a vector observations or measurements, and $\mathbf{x}$ is a vector parameter [7], and the aim is to estimate $\mathbf{x}$ with the vector observations to have minimum error. In other words, the answer given in (9) is the linear least squared (LS) estimator of $\mathbf{x}$. Moreover, $(\mathbf{W}^H \mathbf{W})^{-1} \mathbf{W}^H$ is called a pseudo inverse of $\mathbf{W}$. Since the matrix $\mathbf{W}$ has orthogonal columns, $\mathbf{W}^H \mathbf{W} = K\mathbf{I}_N$ is always satisfied ($\mathbf{I}_N$ is the identity matrix of size $N$). It follows that

$$(\mathbf{W}^H \mathbf{W})^{-1} = (1/K)\mathbf{I}_N \quad \rightarrow \quad \hat{\mathbf{x}} = (1/K)\mathbf{W}^H \mathbf{y_m} \quad (10)$$

where $\hat{\mathbf{x}}$ is linear estimator of vector parameter $\mathbf{x}$. Since the amplitude of the desired signal is constant, $\bar{\hat{\mathbf{x}}}$ is defined as vector $\hat{\mathbf{x}}$ with normalized coefficients. We can processed to calculate minimum error in below.

$$Error_{Min} = \|\mathbf{y_m} - \mathbf{W}\bar{\hat{\mathbf{x}}}\|^2 \quad (11)$$

where $\|.\|$ is $l^2$-norm on vector space. The vector $\mathbf{y_m}$ is constructed by the phase values of $\theta(k)$ which should be improved at each iteration; therefore, algorithm is designed in a way that $\mathbf{y_m}$ and then $\bar{\hat{\mathbf{x}}}$ are obtained proportional to $\theta(k)$. Now, $\bar{\hat{\mathbf{x}}}$ can be given for r-th iteration as shown below.

$$\bar{\hat{\mathbf{x}}}^{(r)} = \exp\left(j\text{phase}\left(\mathbf{W}^H \mathbf{y_m}^{(r-1)}\right)\right) \quad (12)$$

Having determined $\bar{\hat{\mathbf{x}}}^{(r)}$, we can calculate $\boldsymbol{\theta}$ in the iteration of r-th by taking DFT of $\bar{\hat{\mathbf{x}}}^{(r)}$

$$\boldsymbol{\theta}^{(r)} = \text{phase}\left(\mathbf{W}\bar{\hat{\mathbf{x}}}^{(r)}\right) \quad (13)$$

then $\mathbf{y_m^{(r)}}$ will be obtained as follow

$$\mathbf{y_m^{(r)}} = \begin{bmatrix} |Y(0)|\exp\left(j\theta^{(r)}(0)\right) \\ |Y(1)|\exp\left(j\theta^{(r)}(1)\right) \\ |Y(2)|\exp\left(j\theta^{(r)}(2)\right) \\ \vdots \\ |Y(K-1)|\exp\left(j\theta^{(r)}(K-1)\right) \end{bmatrix} \quad (14)$$

At the end of r-th iteration, designed signal is $\bar{\bar{\mathbf{x}}}^{(r)}$, that can be multiplied by the constant coefficient $A$. We shall write the minimum error of the r-th iteration as follow

$$Error_{Min}^{(r)} = \left\| \mathbf{y_m^{(r-1)}} - \mathbf{W}\bar{\bar{\mathbf{x}}}^{(r)} \right\|^2 \quad (15)$$

Noteworthy, to start the algorithm from an appropriate point, the obtained phase from the stationary phase concept [8] is taken as the initial phase ($\boldsymbol{\theta}^{(0)}$), then, by following the algorithm, we try to get close to the optimal condition. The proposed algorithm is efficient as long as the minimum error value is reducing in each iteration. In the other words, algorithm holds to be efficient when the following relation is true

$$\Delta Error_{Min}^{(r)} = Error_{Min}^{(r)} - Error_{Min}^{(r-1)} < 0 \quad (16)$$

As long as (16) is satisfied, the proposed phase improvement algorithm will be efficient. In next section, the minimum error is discussed with regard to the number of iterations to analyze the algorithm efficiency.

*Simulation and Results:* The phase improvement algorithm is performed for the four windows of Raised-Cosine, Taylor, Chebyshev, and Kaiser. Design parameters such as pulse width, bandwidth, and sampling rate are considered as 2.5 $\mu s$, 100 MHz and 1 GHz, respectively. Fig. 1 illustrates the one-sided autocorrelation function of the designed signal using the phase improvement algorithm for the four windows of Raised-Cosine, Taylor, Chebyshev, and Kaiser.

Table 1 compares the results of the phase improvement algorithm with the stationary phase method for the peak sidelobe level (PSL) of the autocorrelation function. According to Table 1, the average reduction of PSL in the phase improvement algorithm is about 5 $dB$ compared with the stationary phase method. Additionally, The PSL reduction by the Kaiser window is greater than the other windows indicating its superior performance in this algorithm.

The minimum error of the four windows of Raised-Cosine, Taylor, Chebyshev, and Kaiser is also calculated with regard to (15) for each iteration, and is shown in Fig. 2. The results demonstrate that the algorithm initially hold a significant error, but it tends to decrease and in high number of iteration it has an almost constant variation. This minimum error variations provide the condition obtained in (16) for the efficiency of the proposed phase improvement algorithm.

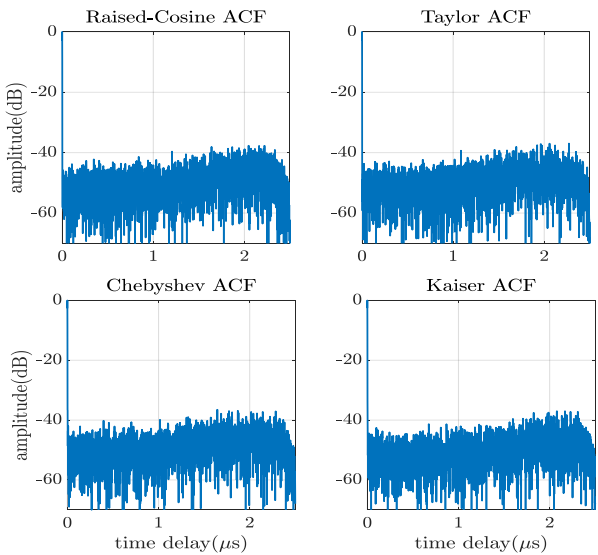

**Fig. 1** *The one-sided autocorrelation function of the designed signal using the phase improvement algorithm for the four windows of Raised-Cosine, Taylor, Chebyshev, and Kaiser.*

**Table 1:** PSL Comparison for the Phase Improvement Algorithm and Stationary Phase Method.

| Windows | PSL with SPC (dB) [8] | PSL with PIA (dB) | Improvement (dB) |
|---|---|---|---|
| Raised-Cosine | -33.34 | -37.67 | -4.33 |
| Taylor | -33.34 | -37.38 | -4.04 |
| Chebyshev | -31.77 | -36.50 | -4.73 |
| Kaiser | -30.98 | -36.98 | -6.00 |

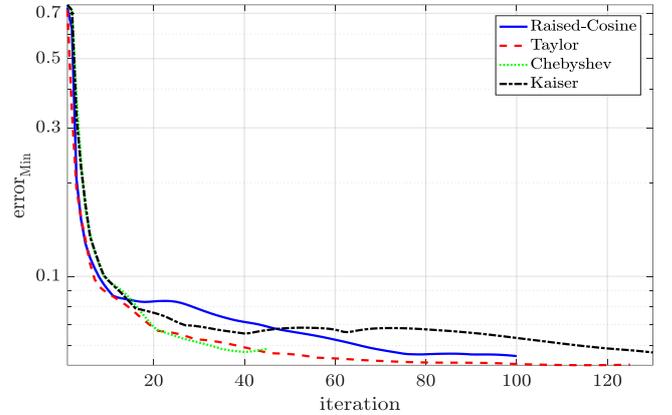

**Fig. 2** *Minimum error against the number of iterations for the four windows of Raised-Cosine, Taylor, Chebyshev, and Kaiser.*

*Conclusion:* Phase improvement algorithm demonstrates the reduction of the peak sidelobe level about 5 *dB* with respect to stationary phase method, which is a significant value. The phase improvement for the Kaiser window is higher than the other windows in the proposed algorithm. In fact, the proposed method comes after the stationary phase method for finding an optimal phase. In the design of NLFM signals due to the constant amplitude, the main concern is to find optimal phase. According to the presented results, the minimum error in each iteration has been reduced indicating the superior efficiency of the proposed algorithm.

R. Ghavamirad, M. A. Sebt and H. Babashah (*Department of Electrical Engineering K. N. Toosi University of Technology, Tehran, Iran*)
E-mail: sebt@kntu.ac.ir